\documentstyle[11pt]{article}
\begin{document}
\begin{titlepage}
\hfill{UQMATH-93-08}
\vskip.3in
\begin{center}
{\huge Once More about Spectral-Dependent Quantum $R$-Matrix for
$U_q(A_2)$}
\vskip.3in
{\Large Anthony J. Bracken, Mark D. Gould} and {\Large Yao-Zhong Zhang}
\vskip.3in
{\large Department of Mathematics, University of Queensland, Brisbane,
Qld 4072, Australia}
\end{center}
\vskip.6in
\begin{center}
{\bf Abstract:}
\end{center}
The recently obtained results in \cite{ZG2} are used to compute the
explicitly spectral-dependent quantum $R$-matrix (or the intertwiners) on
$V_{(6)}(x)\otimes V_{(6)}(y)$ and
$V_{(3)}(x)\otimes V_{(6)}(y)$, where $V_{(6)}$ and $V_{(3)}$ are
the 6-dimensional and fundamental representations of $U_q(A_2)$, respectively.
It appears that the $R$-matrix on $V_{(3)}(x)\otimes V_{(6)}(y)$
depends on $q$ in the different way from what one might usually think:
$q$ occurs in the $R$-matrix in fractional powers.  It seems to be
the first example in literatures of $R$-matrix with the new feature.

\end{titlepage}

\noindent In our previous papers \cite{ZG2} we determine the explicitly
spectral-dependent universal $R$-matrix for the quantum algebras $U_q(A_1)$
and $U_q(A_2)$ by using the universal $R$-matrix \cite{KT}\cite{ZG1}
for the corresponding quantum
affine algebras $U_q(A_1^{(1)})$ and $U_q(A_2^{(1)})$, respectively;
applying the results to the adjoint representation of $U_q(A_2)$,
the simplest non-trivial case where the tensor product of the representation
with itself is with finite
multiplicity, we obtain the explicitly spectral-dependent solution to the
quantum Yang-Baxter equation \cite{Drinfeld}\cite{Jimbo}\cite{Reshetikhin}
associated to the the adjoint representation of $U_q(A_2)$.

The present letter continues the above investigations to other interesting
cases to find the spectral-dependent quantum $R$-matrix on
modules $V_{(6)}(x)\otimes V_{(6)}(y)$
and $V_{(3)}(x)\otimes V_{(6)}(y)$, where $V_{(6)}$ and $V_{(3)}$ are the
6-dimensional and fundamental representations of $U_q(A_2)$, respectively.

The former may be in principal treated by means of the usual
"Yang-Baxterization" procedure used in literatures
\cite{Jones}\cite{Jimbo}\cite{ZGB}\cite{MacKay},
however, trying to find an explicit expression for the $R$-matrix
by this procedure is inevitably involved in the
determination of the explicit CG coefficients for $U_q(A_2)$
which as is well-known is very complicated and is not generally avaiable
even for the corresponding classical algebra $A_2$. The latter
could not be treated by the "Yang-Baxterization" approach since braid
group generators always act on the tensored space of the {\em same}
representation. This may explain why all previous
attempts in literatures at finding spectral-dependent $R$-matrix are only
limitted in the case where the representations being tensored are same. On
the contrast, our approach developed in \cite{ZG2} not only gives rise
to an unified treatment for the case of whether the tensor product is
multiplicity-free or with finite mulltiplicity, as showed in \cite{ZG2},
but also to an unified treatment for the case of whether the representations
being tensored are same or different, as we will see below.

There is more to that. As one may see in (\ref{36R}) below, the $R$-matrix
or the intertwiner on $V_{(3)}(x)\otimes V_{(6)}(y)$ behaves different from
what one might usually think: $q$ appears in the $R$-matrix in the form of its
fractional powers. To our knowledge, it is the first example in literatures
of quantum $R$-matrix with the new feature.
Therefore, the present study should throw light on
new kind of solutions to the quantum Yang-Baxter equation.

We begin with a brief review on some results obtained in \cite{ZG2}.
Throughout this letter we use the notations
\begin{eqnarray}
&&(n)_q=\frac{1-q^n}{1-q}\,,~~[n]_q=\frac{q^n-q^{-n}}{q-q^{-1}}\,,~~
  q_\alpha=q^{(\alpha,\alpha)}\nonumber\\
&&{\rm exp}_q(x)=\sum_{n\geq 0}\frac{x^n}{(n)_q!}\,,~~(n)_q!=
  (n)_q(n-1)_q\,...\,(1)_q
\end{eqnarray}
In ref.\cite{ZG2}, starting from the universal $R$-matrix for the quantum
(non-twisted) affine algebra $U_q(A_2^{(1)})$, we determine
the explicitly spectral-dependent universal
$R$-matrix $R(x,y)$ for the corresponding algebra $U_q(A_2)$, given by,
\begin{eqnarray}
R(x,y)&=&\prod_{n\geq 0}~{\rm exp}_{q_\alpha}
\left ((q-q^{-1})\left (\frac{x}{y}\right )^n\left (q^{-nh_\alpha}E_\alpha
q^{-n(h_\alpha+2h_\beta)/3}\otimes q^{n(h_\alpha+2\beta)/3}F_\alpha
q^{nh_\alpha}\right )\right )\nonumber\\
& &\cdot\prod_{n\geq 0}~{\rm exp}_{q_{\alpha+\beta}}
\left ((q-q^{-1})\left (\frac{x}{y}\right )^n\left (q^{-nh_{\alpha+\beta}}
E_{\alpha+\beta}q^{n(h_\beta-h_\alpha)/3}
\otimes q^{n(h_\alpha-\beta)/3}F_{\alpha+\beta}q^{nh_{\alpha+\beta}}
\right )\right )\nonumber\\
& &\cdot \prod_{n\geq 0}~{\rm exp}_{q_\beta}
\left ((q-q^{-1})\left (\frac{x}{y}\right )^n\left (E'_{\beta+n\delta}
\otimes F'_{\beta+n\delta}\right )\right )\nonumber\\
& &\cdot {\rm exp}\left (\sum_{n>0}\sum^2_{i,j=1}
C^n_{ij}(q)(q-q^{-1})\left (\frac{x}{y}\right )^n
(E^{'(i)}_{n\delta}\otimes F^{'(j)}_{n\delta})\right )\nonumber\\
& &\cdot \prod_{n\geq 0}~{\rm exp}_{q_{\beta}}
\left ((q-q^{-1})\left (\frac{x}{y}\right )^{n+1}\left (E'_{(\delta-\beta)
+n\delta}\otimes F'_{(\delta-\beta)+n\delta}\right )\right )\nonumber\\
& &\cdot \prod_{n\geq 0}~{\rm exp}_{q_\alpha}
\left ((q-q^{-1})\left (\frac{x}{y}\right )^{n+1}\left (
q^{-(n+1)(h_\alpha+2h_\beta)/3}F_\alpha q^{-nh_\alpha}
\otimes q^{nh_\alpha}E_\alpha q^{(n+1)(h_\alpha+2h_\beta)/3}
\right )\right )\nonumber\\
& &\cdot\prod_{n\geq 0}~{\rm exp}_{q_{\alpha+\beta}}
\left ((q-q^{-1})\left (\frac{x}{y}\right )^{n+1}\left (q^{(n+1)(h_\beta-
h_\alpha)/3}F_{\alpha+\beta}q^{-nh_{\alpha+\beta}}\right .\right .\nonumber\\
& &\left .\left .~~~~~
\otimes q^{nh_{\alpha+\beta}}E_{\alpha+\beta}q^{(n+1)(h_\alpha-h_\beta)/3}
\right )\right )
\cdot q^{\sum^2_{i,j=1}\,(a^{-1}_{\rm sym})^{ij}h_i\otimes h_j}\label{loop-R}
\end{eqnarray}
where  $\alpha$ and $\beta$ are the simple roots of $A_2$ and
\begin{eqnarray}
&&E_{\alpha+\beta}=E_\alpha E_\beta-q^{-1}E_\beta E_\alpha\,,~~~~
  F_{\alpha+\beta}=F_\beta F_\alpha-q F_\alpha F_\beta\nonumber\\
&&(a_{\rm sym}^{ij})=\left (
\begin{array}{cc}
2 & -1\\
-1 & 2
\end{array} \right )\nonumber\\
&&(C^n_{ij}(q))=(C^n_{ji}(q))=
 \frac{n}{[n]_q}\,\frac{[2]_q^2}{q^{2n}+1+q^{-2n}}\,\left (
\begin{array}{cc}
q^n+q^{-n} & (-1)^n\\
(-1)^n & q^n+q^{-n}
\end{array} \right )\nonumber\\
&&E'_{\beta+n\delta}=(-1)^n[2]_q^{-n}q^n\left \{\left ({\rm ad'}_{q^{-1}
  }{\cal E}\right )^nE_\beta\right \}q^{n(h_\beta-h_\alpha)/3}\nonumber\\
&&F'_{\beta+n\delta}=[2]_q^{-n}q^{n(h_\alpha-h_\beta)/3}
  \left ({\rm ad'}_{q^{-1}}{\cal F}\right )^nF_\beta\nonumber\\
&&E'_{(\delta-\beta)+n\delta}=[2]_q^{-n}q^{-n}\left \{\left ({\rm ad'}_q
  {\cal E}\right )^n({\rm ad'}_{q^{-2}}E_\alpha)F_{\alpha+\beta}
  \right \}q^{(n+1)(h_\beta-h_\alpha)/3}\nonumber\\
&&F'_{(\delta-\beta)+n\delta}=(-1)^n[2]_q^{-n}
  q^{(n+1)(h_\alpha-h_\beta)/3}
  \left ({\rm ad'}_q{\cal F}\right )^n({\rm ad'}_{q^2}
  E_{\alpha+\beta})F_\alpha\nonumber\\
&&{\cal E}=({\rm ad'}_{q^{-1}}E_\beta)({\rm ad'}_{q^{-2}}E_\alpha)
  F_{\alpha+\beta}\,,~~~~{\cal F}=({\rm ad'}_q({\rm ad'}_{q^2}
  E_{\alpha+\beta})F_\alpha)F_\beta\,,
\end{eqnarray}
$({\rm ad'}_Q{\cal A})\cdot {\cal B}\equiv {\cal A}{\cal B}-Q{\cal B}{\cal A}$
{}~;~ $E^{'(i)}_{n\delta}$ and $F^{'(i)}_{n\delta}$ are determined by the
equalities of formal series: ($\alpha_i=\alpha,\,\beta$)
\begin{eqnarray}
&&(q_{\alpha_i}-q_{\alpha_i}^{-1})\sum_{k=1}^\infty\tilde{E}^{'(i)}_
  {k\delta}u^k={\rm exp}\left ((q_{\alpha_i}-q_{\alpha_i}^{-1})
  \sum_{l=1}^\infty E^{'(i)}_{l\delta}u^l\right )-1\nonumber\\
&&-(q_{\alpha_i}-q_{\alpha_i}^{-1})\sum_{k=1}^\infty\tilde{F}^{'(i)}_
  {k\delta}u^{-k}={\rm exp}\left (-(q_{\alpha_i}-q_{\alpha_i}^{-1})
  \sum_{l=1}^\infty F^{'(i)}_{l\delta}u^{-l}\right )-1\label{primed}
\end{eqnarray}
in which
\begin{eqnarray}
&&\tilde{E}^{'(\alpha)}_{n\delta}=(-1)^{n-1}[2]^{-1}_q(E_\alpha F_\alpha
  -q^{-2n}F_\alpha E_\alpha)\,q^{-(n-1)h_\alpha}q^{-n(h_\alpha+2h_\beta)/3}
  \nonumber\\
&&\tilde{F}^{'(\alpha)}_{n\delta}=(-1)^{n-1}[2]^{-1}_q
  \,q^{(n-1)h_\alpha}q^{n(h_\alpha+2h_\beta)/3}(F_\alpha E_\alpha-q^{2n}
  E_\alpha F_\alpha)\nonumber\\
&&\tilde{E}^{'(\beta)}_{n\delta}=(-1)^n[2]_q^{-n}q^{n-2}\left \{
  \left ({\rm ad'}_{q^{-n+2}}{\cal F'}\right )\cdot\left ({\rm ad'}_{q^{-1}}
  {\cal E}\right )^{n-1}E_\beta\right \}q^{n(h_\beta-h_\alpha)/3}\nonumber\\
&&\tilde{F}^{'(\beta)}_{n\delta}=[2]_q^{-n}q^{n-1}q^{n(h_\alpha-h_\beta)/3}
  \left ({\rm ad'}_{q^{-n+2}}{\cal E'}\right )\cdot
  \left ({\rm ad'}_{q^{-1}}{\cal F}\right )^{n-1}F_\beta\nonumber\\
&&{\cal E'}=E_{\alpha+\beta}F_\alpha-q^2F_\alpha E_{\alpha+\beta}\,,~~~~
  {\cal F'}=E_{\alpha}F_{\alpha+\beta}-q^{-2}F_{\alpha+\beta} E_{\alpha}
\end{eqnarray}

It is shown in our previous papers \cite{ZG2} that
the explicit form of generators on the fundamental
representation of $U_q(A_2)$ is given by
\begin{eqnarray}
&&h_\alpha={\rm diag}(1,-1,0)\,,~~~~~~h_\beta={\rm diag}(0,1,-1)\nonumber\\
&&E_\alpha=e_{12}\,,~~~F_\alpha=e_{21}\,,~~~E_{\alpha+\beta}=e_{13}\,,~~~
  F_{\alpha+\beta}=e_{31}\nonumber\\
&&E'_{\beta+n\delta}=q^{-2n-n/3}e_{23}\,,~~~
  F'_{\beta+n\delta}=q^{2n+n/3}e_{32}\nonumber\\
&&E'_{(\delta-\beta)+n\delta}=-q^{-2n-1-(n+1)/3}e_{32}\,,~~~
  F'_{(\delta-\beta)+n\delta}=-q^{2n+1+(n+1)/3}e_{23}\nonumber\\
&&E^{'(\alpha)}_{n\delta}=[2]_q^{-1}(-1)^{n-1}\frac{[n]_q}{n}q^{-n/3}
  \,\left (e_{11}-q^{-2n}e_{22}\right )\nonumber\\
&&F^{'(\alpha)}_{n\delta}=[2]_q^{-1}(-1)^{n-1}\frac{[n]_q}{n}q^{n/3}
  \,\left (e_{11}-q^{2n}e_{22}\right )\nonumber\\
&&E^{'(\beta)}_{n\delta}=-[2]_q^{-1}\frac{[n]_q}{n}q^{-n/3}
  \,\left (q^{-n}e_{22}-q^{-3n}e_{33}\right )\nonumber\\
&&F^{'(\beta)}_{n\delta}=-[2]_q^{-1}\frac{[n]_q}{n}q^{+n/3}
  \,\left (q^ne_{22}-q^{3n}e_{33}\right )\label{act}
\end{eqnarray}
where (and below) $e_{ij}$ is the matrix satisfying $(e_{ij})_{kl}=
\delta_{ik}\delta_{jl}$ and $e_{ij}e_{kl}=\delta_{jk}e_{il}$.
Upon applying (\ref{loop-R}) to $V_{(3)}\otimes V_{(3)}$ in which
$V_{(3)}$ denotes the fundamental representation of $U_q(A_2)$, we get
\cite{ZG2}
\begin{eqnarray}
R_{(3),(3)}(x,y)&=&f_q(x,y)\cdot \left (e_{11}+e_{99}
  +\frac{q^{-1}(y-x)}{y-q^{-2}x}(e_{22}+e_{33}+e_{44}+e_{66}+e_{77}+e_{88})
  +\right .\nonumber\\
& &\left .+\frac{y(1-q^{-2})}{y-q^{-2}x}(e_{24}+e_{37}+e_{68})
  +\frac{x(1-q^{-2})}{y-q^{-2}x}(e_{42}+e_{73}+e_{86})\right )\label{33R}
\end{eqnarray}
where
\begin{equation}
f_q(x,y)=q^{2/3}\cdot {\rm exp}\left (\sum_{n>0}\frac{q^{2n}-q^{-2n}}{q^{2n}+1
 +q^{-2n}}\,\frac{(x/y)^n}{n}\right )\label{scalar}
\end{equation}
Eq.(\ref{33R}) is the well known result,
up to the scalar factor $f_q(x,y)$,
due to Jimbo \cite{Jimbo}. We reproduce it by our approach.

We now come to our main concern in this letter. First we wnat to
apply (\ref{loop-R}) to $V_{(6)}\otimes V_{(6)}$, to find the
quantum $R$-matrix, $R_{(6),(6)}(x,y)$, on the tensor product module
for $U_q(A_2)$.  As we did in \cite{ZG2},
we introduce the so-called Gelfand-Tsetlin basis vector $|(m)>$ given by
\begin{equation}
|(m)>=\left |\left (
\begin{array}{c}
m_{13}~~~m_{23}~~~m_{33}\\
m_{12}~~~m_{22}\\
m_{11}
\end{array}
\right )\right >
\end{equation}
For the 6-dimensional representation, we have the following  $6$ basis vectors:
\begin{eqnarray}
&&\phi_1=\left |\left (
\begin{array}{c}
4/3~~~-2/3~~~-2/3\\
4/3~~~-2/3\\
4/3
\end{array}
\right )\right >\,,~~~\phi_2=\left |\left (
\begin{array}{c}
4/3~~~-2/3~~~-2/3\\
4/3~~~-2/3\\
1/3
\end{array}
\right )\right >\nonumber\\
&&\phi_3=\left |\left (
\begin{array}{c}
4/3~~~-2/3~~~-2/3\\
4/3~~~-2/3\\
-2/3
\end{array}
\right )\right >\,,~~~
\phi_4=\left |\left (
\begin{array}{c}
4/3~~~-2/3~~~-2/3\\
1/3~~~-2/3\\
1/3
\end{array}
\right )\right >\nonumber\\
&&\phi_5=\left |\left (
\begin{array}{c}
4/3~~~-2/3~~~-2/3\\
1/3~~~-2/3\\
-2/3
\end{array}
\right )\right >\,,~~~\phi_6=\left |\left (
\begin{array}{c}
4/3~~~-2/3~~~-2/3\\
-2/3~~~-2/3\\
-2/3
\end{array}
\right )\right >
\end{eqnarray}
Then long computations plus induction in $n$ show that
the matrix form of generators in the 6-dimensional
representation of $U_q(A_2)$ is given by
\begin{eqnarray}
&&h_\alpha={\rm diag}(2,0,-2,1,-1,0)\,,~~~~h_\beta={\rm diag}(
  0,1,2,-1,0,-2)\nonumber\\
&&E_\alpha=[2]_q^{1/2}e_{12}+[2]_q^{1/2}e_{23}+e_{45}\,,~~~~
  F_\alpha=[2]_q^{1/2}e_{21}+[2]_q^{1/2}e_{32}+e_{54}\nonumber\\
&&E_\beta=e_{24}+[2]_q^{1/2}e_{35}+[2]_q^{1/2}e_{56}\,,~~~~
  F_\beta=e_{42}+[2]_q^{1/2}e_{53}+[2]_q^{1/2}e_{65}\nonumber\\
&&E_{\alpha+\beta}=[2]_q^{1/2}e_{14}+qe_{25}+[2]_q^{1/2}e_{46}\,,~~~~
  F_{\alpha+\beta}=[2]_q^{1/2}e_{41}+qe_{52}+[2]_q^{1/2}e_{64}\nonumber\\
&&E'_{\beta+n\delta}=q^{-2n/3}\{q^{-3n}e_{24}+[2]_q^{1/2}q^{-3n}e_{35}+
  [2]_q^{1/2}q^{-n}e_{56}\}\nonumber\\
&&F'_{\beta+n\delta}=q^{2n/3}\{q^{3n}e_{42}+[2]_q^{1/2}q^{3n}e_{53}+
  [2]_q^{1/2}q^{n}e_{65}\}\nonumber\\
&&E'_{(\delta-\beta)+n\delta}=-q^{-2(n+1)/3}\{q^{-3n-2}e_{42}+[2]_q^{1/2}
  q^{-3n-1}e_{53}+[2]_q^{1/2}q^{-n-1}e_{65}\}\nonumber\\
&&F'_{(\delta-\beta)+n\delta}=-q^{2(n+1)/3}\{q^{3n+2}e_{24}+[2]_q^{1/2}q^{3n+1}
  e_{35}+[2]_q^{1/2}q^{n+1}e_{56}\}\nonumber\\
&&E^{'(\alpha)}_{n\delta}=[2]_q^{-1}(-1)^{n-1}\frac{[n]_q}{n}q^{-2n/3}\left \{
  (q^n+q^{-n})e_{11}+q^{-n}(q^{2n}-q^{-2n})e_{22}\right .\nonumber\\
&&~~~~~~~~~~~\left . -q^{-2n}(q^n+q^{-n})e_{33}
  +q^{n}e_{44}-q^{-n})e_{55}\right \}\nonumber\\
&&F^{'(\alpha)}_{n\delta}=[2]_q^{-1}(-1)^{n-1}\frac{[n]_q}{n}q^{2n/3}\left \{
  (q^n+q^{-n})e_{11}-q^{n}(q^{2n}-q^{-2n})e_{22}\right .\nonumber\\
&&~~~~~~~~~~~\left .-q^{2n}(q^n+q^{-n})e_{33}
  +q^{-n}e_{44}-q^{n})e_{55}\right \}\nonumber\\
&&E^{'(\beta)}_{n\delta}=[2]_q^{-1}\frac{[n]_q}{n}q^{-2n/3}\left \{
  -q^{-2n}e_{22}-q^{-n}(q^n+q^{-n})e_{33}+q^{-4n}e_{44}\right .\nonumber\\
&&~~~~~~~~~~~\left .  -q^{-2n}(q^{2n}-q^{-2n})e_{55}+q^{-3n}(q^n
  +q^{-n})e_{66}\right \}\nonumber\\
&&F^{'(\beta)}_{n\delta}=-[2]_q^{-1}\frac{[n]_q}{n}q^{2n/3}\left \{
  -q^{2n}e_{22}-q^{n}(q^n+q^{-n})e_{33}+q^{4n}e_{44}\right .\nonumber\\
&&~~~~~~~~~~~ \left . +q^{2n}(q^{2n}-q^{-2n})e_{55}+q^{3n}(q^n
  +q^{-n})e_{66}\right \}\label{action}
\end{eqnarray}
We have the following properties for the
generators in (\ref{action}),
\begin{eqnarray}
&&(E_\alpha)^2=[2]_qe_{13}\,,~~~(E_\alpha)^3=0\,,~~~(F_\alpha)^2=[2]_qe_{31}
  \,,~~~(F_\alpha)^3=0\nonumber\\
&&(E_{\alpha+\beta})^2=-[2]_qe_{16}\,,~~(E_{\alpha+\beta})^3=0
  \,,~~(F_{\alpha+\beta})^2=-[2]_qe_{61}
  \,,~~(F_{\alpha+\beta})^3=0\nonumber\\
&&(E'_{\beta+n\delta})^2=[2]_qq^{-5n-n/3}e_{36}
  \,,~~(E'_{\beta+n\delta})^3=0\,,~~
  (F'_{\beta+n\delta})^2=[2]_qq^{5n+n/3}e_{63}\,,~~
  (F'_{\beta+n\delta})^3=0\nonumber\\
&&(E'_{(\delta-\beta)+n\delta})^2=[2]_qq^{-5n-3-(n+1)/3}e_{63}\,,~~
  (E'_{(\delta-\beta)+n\delta})^3=0\nonumber\\
&&(F'_{(\delta-\beta)+n\delta})^2=[2]_qq^{5n+3+(n+1)/3}e_{36}\,,~~
  (F'_{(\delta-\beta)+n\delta})^3=0\label{nilpotent}
\end{eqnarray}
as can be easily checked.

Inserting (\ref{action}) into (\ref{loop-R}), we see that in the expansion
of each $q$-exponential only three terms survive thanks to the celebrated
properties of generators, eq.(\ref{nilpotent}). Thus one is able to work
out the infinite products in (\ref{loop-R}).
The contributions from the imaginary root vectors in
(\ref{loop-R}) can also be worked out term by term and written as a very
compact form. The final result may be put in the explicit and
compact form,
\begin{eqnarray}
R_{(6),(6)}(x,y)&=&\left \{1+(q-q^{-1})\sum_{n=0}^\infty\left (\frac{x}{y}
  \right )^n \left (E'_{\alpha+n\delta}\otimes F'_{\alpha+n\delta}\right )
  +y^2f(e_{13}\otimes e_{31})\right \}\nonumber\\
& &\cdot\left \{1+(q-q^{-1})\sum_{n=0}^\infty\left (\frac{x}{y}\right )^n
  \left (E'_{\alpha+\beta+n\delta}\otimes F'_{\alpha+\beta+n\delta}\right )
  +y^2f(e_{16}\otimes e_{61})\right \}\nonumber\\
& &\cdot\left \{1+(q-q^{-1})\sum_{n=0}^\infty\left (\frac{x}{y}\right )^n
  \left (E'_{\beta+n\delta}\otimes F'_{\beta+n\delta}\right )
  +y^2f(e_{36}\otimes e_{63})\right \}\nonumber\\
& &\cdot q^{-1}\,f_q(x,y)\cdot
  \left \{{\rm imaginary~root~vectors~contribution}\right \}\nonumber\\
& &\cdot\left \{1+(q-q^{-1})\sum_{n=0}^\infty\left (\frac{x}{y}\right )^{n+1}
  \left (E'_{(\delta-\beta)+n\delta}\otimes F'_{(\delta-\beta)+n\delta}\right )
  +x^2f(e_{63}\otimes e_{36})\right \}\nonumber\\
& &\cdot\left \{1+(q-q^{-1})\sum_{n=0}^\infty\left (\frac{x}{y}\right )^{n+1}
  \left (E'_{(\delta-\alpha)+n\delta}\otimes F'_{(\delta-\alpha)+n\delta}
  \right )+x^2f(e_{31}\otimes e_{13})\right \}\nonumber\\
& &\cdot\left \{1+(q-q^{-1})\sum_{n=0}^\infty\left (\frac{x}{y}\right )^{n+1}
  \left (E'_{(\delta-\alpha-\beta)+n\delta}\otimes
  F'_{(\delta-\alpha-\beta)+n\delta}\right )
  +x^2f(e_{61}\otimes e_{16})\right \}\nonumber\\
& &\cdot\left \{q^3(e_{11}\otimes e_{11}+e_{33}\otimes e_{33}+e_{66}\otimes
  e_{66})+q(e_{22}\otimes e_{22}+\right .\nonumber\\
&&+e_{44}\otimes e_{44}+e_{55}\otimes e_{55})+
  q(e_{11}\otimes e_{22}+e_{11}\otimes e_{44}+e_{22}\otimes e_{33}+\nonumber\\
& &+\left .e_{33}\otimes e_{55}+e_{44}\otimes e_{66}+e_{55}\otimes e_{66}+
  \{\leftrightarrow\}\right )
  +q^{-1}\left (e_{11}\otimes e_{33}+e_{11}\otimes e_{55}+\right .\nonumber\\
& &+e_{11}\otimes e_{66}+
  e_{22}\otimes e_{66}+e_{33}\otimes e_{44}+e_{33}\otimes e_{66}+
  \{\leftrightarrow\})+\nonumber\\
& &\left .+  \left (e_{22}\otimes e_{44}+e_{22}\otimes e_{55}+
  e_{44}\otimes e_{55}+\{\leftrightarrow\}\right )\right \}\label{88R}
\end{eqnarray}
where $f_q(x,y)$ is given by (\ref{scalar}) and
$"\{\leftrightarrow\}"$ denotes the interchange of the quantities in
the space $X\otimes Y$\,;\,$E'_{\beta+n\delta}$\,,~$F'_{\beta+n\delta}$
\,,~$E'_{(\delta-\beta)+n\delta}$
\,,~$F'_{(\delta-\beta)+n\delta}$ are given in (\ref{action}) and
\begin{eqnarray}
&&f=[2]_q\,q^{-1}\,(q-q^{-1})^2\,\frac{y+q^4x}{(y^2-x^2)(y-q^2x)}\nonumber\\
&&E'_{\alpha+n\delta}=(-1)^nq^{-2n/3}\left \{[2]_q^{1/2}q^{-2n}e_{12}+
  [2]_q^{1/2}e_{23}+e_{45}\right \}\nonumber\\
&&F'_{\alpha+n\delta}=(-1)^nq^{2n/3}\left \{[2]_q^{1/2}q^{2n}e_{21}+
  [2]_q^{1/2}e_{32}+e_{54}\right \}\nonumber\\
&&E'_{\alpha+\beta+n\delta}=(-1)^nq^{-2n/3}\left \{[2]_q^{1/2}q^{-2n}e_{14}+
  qe_{25}+[2]_q^{1/2}e_{46}\right \}\nonumber\\
&&F'_{\alpha+\beta+n\delta}=(-1)^nq^{2n/3}\left \{[2]_q^{1/2}q^{2n}e_{41}+
  qe_{52}+[2]_q^{1/2}e_{64}\right \}\nonumber\\
&&E'_{(\delta-\alpha)+n\delta}=(-1)^nq^{-2(n+1)/3}\left \{[2]_q^{1/2}
  q^{-2n}e_{21}+[2]_q^{1/2}e_{32}+e_{54}\right \}\nonumber\\
&&F'_{(\delta-\alpha)+n\delta}=(-1)^nq^{2(n+1)/3}\left
\{[2]_q^{1/2}q^{2n}e_{12}
  +[2]_q^{1/2}e_{23}+e_{45}\right \}\nonumber\\
&&E'_{(\delta-\alpha-\beta)+n\delta}=(-1)^nq^{-2(n+1)/3}
  \left \{[2]_q^{1/2}q^{-2n}e_{41}+e_{52}+[2]_q^{1/2}e_{64}\right \}\nonumber\\
&&F'_{(\delta-\alpha-\beta)+n\delta}=(-1)^nq^{2(n+1)/3}\left \{[2]_q^{1/2}
  q^{2n}e_{14}+e_{25}+[2]_q^{1/2}e_{46}\right \}\nonumber\\
&&\{{\rm imaginary~root~vectors~contribution}\}=a'/a\,\sum_{i=1}^6\left (
  1+(b/b'-1)(\delta_{i2}+\delta_{i4}+\delta_{i5})\right )(e_{ii}\otimes e_{ii})
  \nonumber\\
&&~~~~~~~~~~~~~~~~+ba'(e_{11}\otimes e_{22}+e_{11}\otimes e_{44})
  +ab(e_{11}\otimes e_{33}
  +e_{11}\otimes e_{55}+e_{11}\otimes e_{66})+\nonumber\\
&&~~~~~~~~~~~~~~~~+ 1/(ab')(e_{22}\otimes e_{11})+a'b(e_{22}\otimes
  e_{33})+ba'/b'(e_{22}\otimes e_{44})+aa'b(e_{22}\otimes e_{55})+\nonumber\\
&&~~~~~~~~~~~~~~~~+ab(e_{22}\otimes e_{66})+1/(a'b')(e_{33}\otimes e_{11})
  +1/(ab')(e_{33}\otimes e_{22})
  +b/b'(e_{33}\otimes e_{44})+\nonumber\\
&&~~~~~~~~~~~~~~~~+ba'(e_{33}\otimes e_{55})+ab(e_{33}\otimes e_{66})+
  1/(ab')(e_{44}\otimes e_{11})+b/(ab')(e_{44}\otimes e_{22})+\nonumber\\
&&~~~~~~~~~~~~~~~~+  b/b'(e_{44}\otimes  e_{33})+
  b/(a'b')(e_{44}\otimes e_{55})+a'b(e_{44}\otimes e_{66})+1/(a'b')
  (e_{55}\otimes e_{11})+\nonumber\\
&&~~~~~~~~~~~~~~~~+1/(aa'b')(e_{55}\otimes e_{22})+1/(ab')(
  e_{55}\otimes e_{33})+b/(ab')(e_{55}\otimes e_{44})+ba'(e_{55}\otimes e_{66})
  +\nonumber\\
&&~~~~~~~~~~~~~~~~+1/(b'a')(e_{66}\otimes e_{11}+
  e_{66}\otimes e_{22}+e_{66}\otimes e_{33})+1/(ab')(e_{66}
  \otimes e_{44})+\nonumber\\
&&~~~~~~~~~~~~~~~~+  1/(ab')(e_{66}\otimes e_{55})\label{generators}
\end{eqnarray}
in which we have defined
\begin{equation}
a=\frac{y-q^2x}{y-x}\,,~~~a'=\frac{y-q^{-2}x}{y-x}\,,~~~b=\frac{y-q^4x}
{y-q^2x}\,,~~~b'=\frac{y-q^{-4}x}{y-q^{-2}x}
\end{equation}
We see that (\ref{88R}) is an extreamly explicit formula: the sums in
(\ref{88R}) can be easily worked out. But for compactness, we leave the
sums in their present form.

We then apply (\ref{loop-R}) to another interesting case:
$V_{(3)}\otimes V_{(6)}$, to derive the quantum $R$-matrix, $R_{(3),(6)}(x,y)$,
on this tensor product module. As we addressed in the introduction,
this case could not treated by means of the
usual Yang-Baxterization procedure simply because the representations being
tensored are different. With the help of (\ref{act}) and (\ref{action})
and through long but similar calculations leading to (\ref{88R}) one finds
\begin{equation}
R_{(3),(6)}(x,y)=g_q(x,y)\cdot R_+\,R_0\,R_-
\end{equation}
where
\begin{eqnarray}
g_q(x,y)&=&q^{1/3}\cdot {\rm exp}\left (\sum_{n>0}\frac{q^{n}-q^{-n}}{q^{2n}+1
  +q^{-2n}}\,\frac{(q^{1/3}\,x/y)^n}{n}\right )\nonumber\\
R_+&=&I+\frac{y(q-q^{-1})}{y-q^{4/3}x}\{[2]_q^{1/2}(e_{12}\otimes e_{21}+
  e_{13}\otimes e_{41}+e_{23}\otimes e_{53})+e_{23}\otimes e_{42}\}\nonumber\\
& &+\frac{y(q-q^{-1})}{y-q^{-2/3}x}\{e_{12}\otimes e_{54}+[2]_q^{1/2}(
  e_{12}\otimes e_{32}+e_{13}\otimes e_{64})+q^{-1}e_{13}\otimes e_{52}+
  \nonumber\\
& &+[2]_q^{1/2}e_{23}\otimes e_{65}\}+\frac{y^2(q-q^{-1})^2}{(y-q^{4/3}x)
  (y-q^{-2/3}x)}e_{13}\otimes e_{52}\nonumber\\
R_0&=&\frac{y-q^{-2/3}x}{y-q^{4/3}x}\{e_{11}\otimes e_{11}+e_{22}\otimes
  e_{33}+e_{33}\otimes e_{66}\}+\nonumber\\
& &\frac{(y-q^{10/3}x)(y-q^{-2/3}x)}{(y-q^{4/3}x)^2}\{e_{11}\otimes e_{22}+
  e_{11}\otimes e_{44}+e_{22}\otimes e_{55}\}\nonumber\\
& &+\frac{y-q^{10/3}x}{y-q^{4/3}x}\{e_{11}\otimes e_{33}+e_{11}\otimes e_{55}
  +e_{11}\otimes e_{66}+e_{22}\otimes e_{66}\}\nonumber\\
& &+\frac{y-q^{-2/3}x}{y-q^{-8/3}x}\{e_{22}\otimes e_{11}+e_{33}\otimes e_{11}
  +e_{33}\otimes e_{22}+e_{33}\otimes e_{33}\}\nonumber\\
& &+\frac{(y-q^{-2/3}x)^2}{(y-q^{4/3}x)(y-q^{-8/3}x)}\{e_{22}\otimes e_{22}
  +e_{33}\otimes e_{44}+e_{33}\otimes e_{55}\}\nonumber\\
& &+\frac{(y-q^{10/3}x)(y-q^{-2/3}x)^2}{(y-q^{4/3}x)^2(y-q^{-8/3}x)}
  e_{22}\otimes e_{44}\nonumber\\
R_-&=&q\{e_{11}\otimes e_{11}+e_{22}\otimes e_{33}+e_{33}\otimes e_{66}\}+
  e_{11}\otimes e_{22}+e_{11}\otimes e_{44}\nonumber\\
& &+e_{22}\otimes e_{22}+e_{22}\otimes e_{55}+e_{33}\otimes e_{44}+
  e_{33}\otimes e_{55}+q^{-1}\{e_{11}\otimes e_{33}+\nonumber\\
& &+e_{11}\otimes e_{55}+e_{11}\otimes e_{66}+e_{22}\otimes e_{11}+e_{22}
  \otimes e_{44}+\nonumber\\
& &+e_{22}\otimes e_{66}+e_{33}\otimes e_{11}+e_{33}\otimes e_{22}+
  +e_{33}\otimes e_{33}+\nonumber\\
& &+\frac{xq^{1/3}(q-q^{-1})}{y-q^{4/3}x}\{[2]_q^{1/2}(e_{21}\otimes e_{12}
  +e_{31}\otimes e_{14}+e_{32}\otimes e_{35})+e_{32}\otimes e_{24}\}
  \nonumber\\
& &+\frac{xq^{-2/3}(q-q^{-1})}{y-q^{-2/3}x}\{[2]_q^{1/2}(e_{21}\otimes e_{23}
  +e_{31}\otimes e_{46})+e_{21}\otimes e_{45}+e_{31}\otimes e_{25}+
  \nonumber\\
& &+[2]_q^{1/2} e_{32}\otimes e_{56}\}+\frac{x^2q^{2/3}(q-q^{-1})^2}
  {(y-q^{4/3}x)(y-q^{-2/3}x)}e_{31}\otimes e_{25}\label{36R}
\end{eqnarray}
It is understood that $I$ in $R_+$ is a $18\times 18$ unit matrix.

To summarize, we have obtained the explicit and compact formulae for
the quantum $R$-matrix (or the intertwiners)
on $V_{(6)}(x)\otimes V_{(6)}(y)$ and $V_{(3)}(x)\otimes V_{(6)}(y)$ of
$U_q(A_2)$. In the latter case, an interesting phenomenon appears:
the powers of $q$ are no longer integers rather fractional numbers!
We believe that it is the first example in the literature
of quantum $R$-matrix with the feature.
The possiblly physical meanings of this new phenomenon remain to be seen.

\vskip.3in
\noindent {\bf Acknowledgements:} Y.Z.Z. would like to thank N.J.MacKay for
e-mail comments and drawing his attention to ref.\cite{MacKay}.
The financial support from Australian Research Council
is gratefully acknowledged.


\end{document}